**Phenotypic divergence of *Homo sapiens* is driven by the evolution of human-specific genomic regulatory networks via two mechanistically distinct pathways of creation of divergent regulatory DNA sequences**

Gennadi V. Glinsky[1]

[1] Institute of Engineering in Medicine

University of California, San Diego

9500 Gilman Dr. MC 0435

La Jolla, CA 92093-0435, USA

Correspondence: gglinskii@ucsd.edu

Web: http://iem.ucsd.edu/people/profiles/guennadi-v-glinskii.html

**Running title:** Divergence pathways of human-specific regulatory sequences

**Abstract**

Thousands of candidate human-specific regulatory sequences (HSRS) have been identified, supporting the hypothesis that unique to human phenotypes result from human-specific alterations of genomic regulatory networks. Here, conservation patterns analysis of 18,364 candidate HSRS was carried out based on definition of the sequence conservation threshold as the minimum ratio of bases that must remap of 1.00. A total of 5,535 candidate HSRS were identified that are: i) highly conserved in Great Apes; ii) evolved by the exaptation of highly conserved ancestral DNA; iii) defined by either the acceleration of mutation rates on the human lineage or the functional divergence from nonhuman primates. The exaptation of highly conserved ancestral DNA pathway seems mechanistically distinct from the evolution of regulatory DNA segments driven by the species-specific expansion of transposable elements. Present analysis supports the idea that phenotypic divergence of *Homo sapiens* is driven by the evolution of human-specific genomic regulatory networks via two mechanistically distinct pathways of creation of divergent sequences of regulatory DNA: i) exaptation of the highly conserved ancestral regulatory DNA segments; ii) human-specific insertions of transposable elements.

**Introduction**

Extensive search for human-specific genomic regulatory sequences (HSRS) revealed thousands candidate HSRS, a vast majority of which is residing within non-protein coding genomic regions (McLean et al., 2011; Shulha et al., 2012; Konopka et al., 2012; Capra et al., 2013; Marnetto et al., 2014; Glinsky, 2015). Candidate HSRS comprise multiple distinct families of genomic regulatory elements, which were defined using a multitude of structural features, different statistical algorithms, as well as a broad spectrum of experimental, analytical, computational, and bioinformatics strategies. The current catalogue of candidate HSRS includes conserved in humans novel regulatory DNA sequences designated human accelerated regions, HARs (Capra et al., 2013); fixed human-specific regulatory regions, FHSRR (Marnetto et al., 2014); human-specific transcription factor-binding sites, HSTFBS (Glinsky, 2015), regions of human-specific loss of conserved regulatory DNA termed hCONDEL (McLean et al., 2011); human-specific epigenetic regulatory marks consisting of H3K4me3 histone methylation signatures at transcription start sites in prefrontal neurons (Shulha et al., 2012); and human-specific transcriptional genetic networks in the frontal lobe (Konopka et al., 2012). Most recently, Gittelman et al. (2015) reported identification of 524 DNase I hypersensitive sites (DHSs) that are conserved in nonhuman primates but accelerated in the human lineage (haDHS) and may have contributed to human-specific phenotypes. They estimated that 70% of substitutions in haDHSs are attributable to positive selection consistent with the hypothesis that these DNA segments have been subjects to human-specific adaptive evolution resulting in creation of human-specific regulatory sequences. Finally, Prescott et al. (2015) identified thousands of enhancers associated with divergent cis-regulatory evolution of the human's and chimpanzee's neural crest underlying development of unique to human craniofacial features.

Definition of HARs, which is one of the most actively investigated HSRS families, is based on calculations as a baseline the evolutionary expected rate of base substitutions derived from the experimentally determined level of conservation between multiple species at the given locus. The statistical significance of differences between the observed substitution rates on a lineage of interest in relation to the evolutionary expected baseline rate of substitutions can be estimated. This method is considered particularly effective for identifying highly conserved sequences within noncoding genomic regions that have experienced a marked increase of substitution rates on a particular lineage. It has been successfully applied to humans (Pollard et al. 2006; Prabhakar et al. 2006; Bird et al. 2007), where the rapidly-evolving sequences that are highly conserved across mammals and have acquired many sequence changes in humans since divergence from chimpanzees were designated as human accelerated regions (HARs). Experimental analyses of HARs bioactivity revealed that some HARs function as non-coding RNA genes expressed during the neocortex development (Pollard et al., 2006) and human-specific developmental enhancers (Prabhakar et al. 2008). Consistent with the hypothesis that HARs function in human cells as regulatory sequences, most recent computational analyses and transgenic mouse experiments demonstrated that many HARs represent developmental enhancers (Capra et al., 2013).

In contrast to the cross-species quantitative analyses of the DNA sequence conservation and divergence, an alternative approach to discovery of candidate HSRS is based on identification of regulatory DNA segments that are functionally divergent in humans compared with our closest evolutionary relatives, chimpanzee and bonobo (Shulha et al., 2012; Prescott at al., 2015). The systematic analysis of the sequence conservation patterns of these families of candidate HSRS, which were defined based on the functional divergence from the NHP, has not been performed.

Here, the sequence conservation patterns' analyses of 18,364 candidate HSRS was carried out using the most recent releases of reference genomes' databases of humans and nonhuman primates and defining the sequence conservation threshold as the minimum ratio of bases that must remap of 1.00. This analysis identifies 5,535 regulatory DNA segments that are: i) predominantly located within the non-coding genomic regions; ii) highly conserved in humans and other Great Apes; iii) do not intersect transposable elements (TE) - derived sequences; iv) appear to acquire human-specific regulatory traits by exaptation of ancestral DNA. In contrast to the exaptation pathway of the human regulatory DNA divergence, a majority of candidate HSRS intersect TE-derived sequences and appear seeded by TE-associated pathway of the human regulatory DNA evolution. The results of the present analyses suggest that evolution of human-specific genomic regulatory networks is driven by at least two mechanistically distinct pathways of creation of divergent regulatory DNA segments.

**Results and Discussion**

**Effects of the human reference database refinements on the validity of molecular definitions of 18,364 candidate human-specific regulatory sequences**

The sequence quality of reference genome databases is essential for the accurate definition of regulatory DNA segments as candidate HSRS. It was unclear how continuing database improvements would affect the validity of the HSRS' definition. To address this problem, the most recent hg38 release of the human genome reference database (HGRD), which replaces the hg19 release as default human assembly (http://genome.ucsc.edu/cgi-bin/hgGateway), was utilized. Present analyses revealed variable effects of the HGRD refinement's on the validity of molecular definitions of distinct families of candidate HSRS (Tables 1-11). The large HGRD refinements' effect was observed on the molecular definition of 583 hCONDELs (McLean et al., 2011), indicating that only 42% of the hCONDELs' sequences, which were originally defined using the hg18 release of the HGRD, could be mapped to the most recent hg38 release of the HGRD (Table 11). A moderate HGRD refinements' effect was observed on the molecular definition of human-specific epigenetic regulatory sequences consisting of H3K4me3 histone methylation signatures at transcription start sites (TSS) in prefrontal neurons (Shulha et al., 2012), indicating that 16 (3.9%) of 410 H3K4me3 marks defined as candidate HSRS failed to convert to the hg38 release of the HGRD at MinMatch threshold of 1.00 (Table 9). However, in most instances, the required adjustments were limited to a few sequences, thus validating the overall high sequence quality of candidate HSRS.

**Sequence conservation analysis of human accelerated DNase I hypersensitive sites**

The identified haDHSs represent relatively short DNA segments of the median size 290 bp. (range from 150-1010 bp.; average size of 323 bp.), which are predominantly located within intronic and intergenic sequences (Gittelman et al., 2015). To test whether reported 524 haDHSs represent human-specific DNA sequences, the conservation analysis was carried-out using the LiftOver algorithm and Multiz Alignments of 20 mammals (17 primates) of the UCSC Genome Browser on Human Dec. 2013 (GRCh38/hg38) Assembly

(http://genome.ucsc.edu/cgi-bin/hgTracks?db=hg38&position=chr1%3A90820922-90821071&hgsid=441235989_eelAivpkubSY2AxzLhSXKL5ut7TN ).

The most recent releases of the corresponding reference genome databases were utilized to ensure the use of the most precise, accurate, and reproducible genomic DNA sequences available to date. The results of these analyses are reported in the Table 1. Several thresholds of the LiftOver algorithm MinMatch function (minimum ratio of bases that must remap) were utilized to assess the sequences conservation and identify candidate human-specific regulatory sequences as previously described (Glinsky, 2015). In this analysis, the candidate human-specific regulatory sequences were defined based on conversion failures to both Chimpanzee's and Bonobo's genomes and supported by direct visual evidence of human-specific sequence alignment differences of the Multiz Alignments of 20 mammals (17 primates). It appears that only small fractions (0.2%-13.9%) of reported 524 haDHSs can be defined as candidate human-specific regulatory sequences applying these criteria at different sequence conservation thresholds (Table 1). Based on this analysis, the vast majority (86.1% to 99.8%) of 524 haDHSs could be classified as the candidate regulatory sequences that appear conserved in humans and nonhuman primates.

Interestingly, the Multiz Alignments of 20 mammals (17 primates) revealed that 71% of candidate human-specific haDHSs defined at 0.99 MinMatch threshold (Table 1) contain small human-specific inserts of 2-15 bp., suggesting a common mutation mechanism (Supplemental Table S1). A majority (78%) of candidate human-specific haDHSs are located within the intronic (47.9%) and intergenic (30.1%) sequences (Supplemental Table S2). However, 15 of 73 (20.5%) candidate human-specific haDHSs sequences appear to intersect exons, 11 of which include intron/exon junctions (Supplemental Tables S1 & S2). Intriguingly, this analysis identified the 18 bp. human-specific deletion within the exon 9 of the *PAX8* gene, which appears to affect the structure of the *PAX8-AS1* RNA as well (Supplemental Table S1).

Therefore, these analyses demonstrate that there is no detectable reference genome database refinements' effect on the accuracy of molecular definition of haDHSs and the majority of haDHSs' sequences are conserved in humans and nonhuman primates.

**Sequence conservation analysis of human accelerated regions**

Strikingly similar results were observed when the sequence conservation analysis of 2,745 HARs was performed (Table 2). It appears that only small fractions (1.2%-9.3%) of reported HARs can be defined as candidate human-specific regulatory sequences using different sequence conservation thresholds (Table 2). Based on this analysis, the vast majority (90.7% to 98.8%) of 2,745 HARs could be classified as the candidate regulatory sequences that appear conserved in humans and nonhuman primates (Table 2). This conclusion remains valid when the most stringent definition of the sequence conservation threshold was used by setting the minimum sequence alignments' match requirement (MinMatch threshold) as the ratio of bases that must remap of 1.00 (Table 2). Based on this analysis, it appears that there is a minor reference genome database refinements' effect on the accuracy of molecular definition of HARs and the majority of HARs' sequences are conserved in humans and nonhuman primates.

**Sequence conservation analysis of other classes of candidate HSRS**

In contrast to haDHS and HARs, several other classes of candidate HSRS were defined based on the failure of alignments of human regulatory DNA segments to the reference genome databases of other species (Marnetto et al., 2014; Glinsky, 2015). It appears that a majority (82.1%-88.4%) of reported DNase I hypersensitive sites-derived fixed human specific regulatory regions (DHS_FHSRR) can be defined as candidate human-specific regulatory sequences using different sequence conservation thresholds (Table 3). Based on this analysis, the relatively minor fraction (11.6% to 17.9%) of 2,118 DHS_FHSRR may be classified as the candidate regulatory sequences that appear conserved in humans and nonhuman primates (Table 3).

Similarly, a majority (79.0%-86.5%) of reported HSTFBS can be defined as candidate human-specific regulatory sequences using different sequence conservation thresholds (Table 4). The relatively minor fraction (13.5% to 21.0%) of 3,803 HSTFBS may be classified as the candidate regulatory sequences that appear conserved in humans and nonhuman primates (Table 4). A majority (70.2%-79.7%) of reported hESC_FHSRR can be defined as candidate human-specific regulatory sequences using different sequence conservation thresholds (Table 5). The relatively small fraction (20.3% to 29.8%) of 1,932 hESC_FHSRR could be classified as the candidate regulatory sequences that appear conserved in humans and nonhuman primates (Table 6). A majority (84.3%-89.7%) of reported other_FHSRR can be defined as candidate human-specific regulatory sequences using different sequence conservation thresholds (Table 5). The relatively minor fraction (10.3% to 15.7%) of 4,249 other_FHSRR could be classified as the candidate regulatory sequences that appear conserved in humans and nonhuman primates (Table 6). Based on this analysis, it appears that there is a minor reference genome database refinements' effect on the accuracy of molecular definition of HSTFBS and FHSRR families of candidate HSRS. The majority of HSTFBS and FHSRR sequences failed to align to both Chimpanzee and Bonobo genomes, thus meeting the criteria for definition as candidate HSRS.

**Identification of highly conserved in nonhuman primates regulatory DNA sequences among candidate HSRS**

To identify regulatory DNA segments that are highly conserved in nonhuman primates, the most stringent definition of the sequence conservation threshold was used by setting the minimum sequence alignments' match requirement as the ratio of bases that must remap of 1.00. It has been noted that a direct lift over at MinMatch 1.00 from human's genome to genomes of nonhuman primates may identify the aligned sequences with clearly visible base differences detectable during the visual inspections of aligned sequences, which was most often due to the losses of the ancestral DNA. To address this limitation, in the subsequent analysis a given regulatory DNA segment was defined as highly conserved only when both direct and reciprocal conversions between humans' and nonhuman primates' genomes were observed using the MinMatch threshold of 1.00. This approach removed sequences with the ancestral DNA losses during the reciprocal alignments of the corresponding genomes of nonhuman primates to the human reference genome. Nevertheless, the majority of both haDHSs (404 of 524; 77.1%) and HARs (2,262 of 2,739; 82.6%) were defined as the highly conserved in humans and nonhuman primates regulatory sequences (Table 7). In contrast, only relatively small fractions of other classes of candidate HSRS were identified as highly conserved in nonhuman primates regulatory sequences, scoring at 7.3% for HSTFBS; 8.3% for other_FHSRR; 9.4% for DHS_FHSRR; and 15.9% for hESC_FHSRR (Table 7). Follow-up visual inspections of these highly conserved in nonhuman primates' genomes candidate regulatory sequences and nucleotide BLAST analyses of selected sequences revealed examples of the overall similar sequence gap structures among the Great Apes after the divergence from the *Rhesus Macaque*, however, some Great Apes display the unique structure of the sequence gaps for individual species.

Significantly, during the BLAST analyses of these DNA segments the consistently high levels of the sequence identities among different species of primates were observed, ranging from 91% to 100% (Supplemental Tables S3 and S4). Taken into consideration that a majority of haDHS and HARs are located within intronic and intergenic regions, it seems reasonable to conclude that these sequences manifest a high level of sequence conservation in nonhuman primates.

Notably, despite the setting of the MinMatch lift over threshold at 1.00, the follow-up BLAST analyses of selected sequences revealed that humans and Great Apes manifest clearly discernable species-specific patterns of single-nucleotide substitutions (Supplemental Tables S3 and S4). Specifically, this pattern was noted during the BLAST analyses of human, *Chimpanzee*, and *Bonobo* sequences. It is possible that these species-specific single-nucleotide substitutions may be of functional significance. Lastly, it has been confirmed during the present analysis that haDHS sequences display rates of mutations accelerated by 1.7- to 8.0-fold in humans compared with *Bonobo* and *Chimpanzee* genomes (Supplemental Table S5). Calculations of the increased mutation rates within human's and primate's lineages were made based on direct measurements of the sequence identities after the split with the *Gorilla gorilla* ~17 million years ago (Supplemental Table S5). Interestingly, a sub-set of haDHS appears to remain 100% identical in both *Bonobo's* and *Chimpanzee's* genomes during ~25-30 million years of evolution after the split with the *Rhesus Macaque* and undergoes single-nucleotide substitutions in the human lineage after the split with the *Chimpanzee* ~13 million years ago. Examples of these haDHS sequences are shown in the Supplemental Tables S4 and S5.

**Sequence conservation patterns' analyses of candidate HSRS defined by the functional divergence in humans compared with chimpanzees**

It was of interest to analyze the sequence conservation patterns among the candidate HSRS, which were defined based on identification of regulatory DNA segments that are functionally divergent in humans compared with our closest evolutionary relatives, chimpanzee and bonobo (Shulha et al., 2012; Prescott at al., 2015). The results of these analyses recapitulate two major patterns of sequence conservations observed for other families of candidate HSRS (Tables 8-12). The sequence conservation patterns of both human-biased and chimp-biased CNCCs' enhancers resemble the sequence conservation profiles of haDHSs and HARs with the majority of regulatory DNA segments (80.7% and 82.2% for human-biased and chimp-biased CNCCs enhancers, respectively) being defined as highly conserved in human, Bonobo, and Chimpanzee genomes (compare data in Tables 1; 2; 7; and Tables 8; 9; 12). In contrast, human-specific regulatory sequences consisting of H3K4me3 histone methylation signatures at transcription start sites in prefrontal neurons manifest sequence conservation patterns similar to the sequence conservation profiles of the FHSRR and HSTFBS with only the minor fraction of regulatory DNA sequences (12.7%) being identified as highly conserved in human, Bonobo, and Chimpanzee genomes (compare data in the Table 3-7 and Tables 10; 12).

In total, 5,535 candidate HSRS, which were defined by either the acceleration of mutation rates on the human lineage or the functional divergence from chimpanzee, appear highly conserved in humans and NHP. Nonetheless, these sequences manifest clearly discernable species-specific patterns of single-nucleotide substitutions in humans, chimpanzee, and bonobo genomes suggesting that they evolved by the exaptation of ancestral regulatory DNA.

**Conclusions**

The results of the present analyses have important implications for our understanding of mechanisms of biogenesis and evolution of the majority of HARs and haDHS as the candidate HSRS. Based on the sequence conservation analyses using the most recent releases of the reference genome databases, it is proposed to define these predominantly intronic and intergenic DNA segments manifesting more than 90% sequence identities among the Great Apes as the candidate regulatory sequences that are highly conserved in both human and NHP lineages. Using this approach, a total of 5,535 regulatory DNA segments (Supplemental Data Set 1) are classified as the highly conserved in humans and nonhuman primates regulatory sequences, suggesting that these candidate HSRS evolved by the exaptation pathway of ancestral regulatory DNA segments, which is mechanistically distinct from the evolution of regulatory DNA driven by the species-specific expansion of transposable elements. Consistent with this notion, it has been demonstrated that transposable element-derived sequences, most notably LTR7/HERV-H, LTR5_Hs, and L1HS, harbor 99.8% of the candidate human-specific regulatory loci with putative transcription factor-binding sites in the genome of hESC (4).

Present analysis revealed a variable reference database refinement's effect on the validity of molecular definitions of different families of candidate HSRS. It identifies limitations of the current computational cross-species sequence alignment algorithm and underscores the requirement of the careful follow-up analyses of each individual candidate HSRS using the most recent releases of the reference genome databases of Great Apes and other nonhuman primates. A large fraction of regulatory DNA segments representing candidate HSRS appears highly conserved in humans and other Great Apes. Reported herein sequence conservation analysis reveals that a significant majority of haDHSs, HARs, and CNCCs' enhancers appears to represent highly conserved in humans and nonhuman primates candidate regulatory sequences that are consistently manifest species-specific patterns of single-nucleotide substitutions and accelerated mutation rates on the human lineage. Collectively, these observations imply that human-specific phenotypes may evolve as a result of combinatorial interplay of both conserved in nonhuman primates and human-specific (unique to humans) regulatory sequences. Based on the present analyses, it seems reasonable to propose that at least two mechanistically distinct pathways of creation of divergent sequences of regulatory DNA drive the evolution of human-specific regulatory networks (**Figure 1**). Diverse families of candidate HSRS, which were defined by either the acceleration of mutation rates on the human lineage or the functional divergence from chimpanzee, appear highly conserved in humans and NHP, strongly arguing that they evolved via the exaptation of ancestral regulatory DNA. This conclusion is in agreement with recent reports describing exaptation of ancestral DNA as a mechanism of creation of human-specific enhancers active in embryonic limb (Cotney et al., 2013) and as a prevalent mechanism of recently evolved enhancers' creation during the mammalian genome evolution (Villar et al., 2015). Despite the exceedingly high interspecies sequence identities for non-coding genomic regions and only minor differences of DNA sequences estimated in the range of ~3-6 substitutions per 500 bp of the regulatory sequence (Prescott et al., 2015), it appears that the acquisition of a small number of mutations was sufficient to confer biologically discernable divergence of regulatory activities.

**Methods**

**Data source**

Solely publicly available datasets and resources were used for this analysis. A total of 18,364 candidate HSRS were analyzed in this study, including 2,745 human accelerated regions (Capra et al., 2013); 524 human accelerated DNase I hypersensitive sites (Gittelman et al., 2015); 3,083 human-specific transcription factor binding sites (Glinsky, 2015); 8,229 fixed human-specific regulatory regions, FHSRR (Marnetto et al., 2014), which were divided into 2,118 DHS_FHSRR; 1,932 hESC_FHSRR; and 4,249 FHSRR identified in different human cell lines, excluding hESC (Other_FHSRR); 583 regions of human-specific loss of conserved regulatory DNA termed hCONDELs (McLean et al., 2011); 410 human-specific epigenetic regulatory marks consisting of H3K4me3 histone methylation signatures at transcription start sites in prefrontal neurons (Shulha et al., 2012); 1,000 human-biased and 1,000 chimp-biased cranial neural crest cells (CNCC) enhancers, which are associated with divergent cis-regulatory evolution of the human's and chimpanzee's neural crest and development of unique to human craniofacial features (Prescott et al., 2015).

**Data analysis**

To determine the conservation patterns of reported 18,364 candidate human-specific regulatory DNA sequences, the conservation analysis was carried-out using the LiftOver algorithm and Multiz Alignments of 20 mammals (17 primates) of the UCSC Genome Browser (Kent et al., 2002) on Human Dec. 2013 Assembly (GRCh38/hg38) (http://genome.ucsc.edu/cgi-bin/hgTracks?db=hg38&position=chr1%3A90820922-90821071&hgsid=441235989_eeIAivpkubSY2AxzLhSXKL5ut7TN ).

The most recent releases of the corresponding reference genome databases were utilized to ensure the use of the most precise, accurate, and reproducible genomic DNA sequences available to date. A candidate HSRS was considered conserved if it could be aligned to either one or both *Chimpanzee* or *Bonobo* genomes using defined sequence conservation thresholds of the LiftOver algorithm MinMatch function. LiftOver conversion of the coordinates of human blocks to non-human genomes using chain files of pre-computed whole-genome BLASTZ alignments with a specified MinMatch levels and other search parameters in default setting

([http://genome.ucsc.edu/cgi-bin/hgLiftOver](http://genome.ucsc.edu/cgi-bin/hgLiftOver)). Several thresholds of the LiftOver algorithm MinMatch function (minimum ratio of bases that must remap) were utilized to assess the sequences conservation and identify candidate human-specific (MinMatch of 0.95; 0.99; and 1.00) and conserved in nonhuman primates (MinMatch of 1.00) regulatory sequences as previously described (Glinsky, 2015). The Net alignments provided by the UCSC Genome Browser were utilized to compare the sequences in the human genome (hg38) with the mouse (mm10), *Chimpanzee* (PanTro4), and *Bonobo* genomes. A given regulatory DNA segment was defined as the highly conserved regulatory sequence when both direct and reciprocal conversions between humans' and nonhuman primates' genomes were observed using the MinMatch sequence alignment threshold of 1.00 (Tables 7 and 12). A given regulatory DNA segment was defined as the candidate human-specific regulatory sequence when sequence alignments failed to both *Chimpanzee* and *Bonobo* genomes using the specified MinMatch sequence alignment thresholds (Tables 1-6; 8-11).

## Supplemental Information

Supplemental information includes Supplemental Tables S1-S5; Supplemental Data Set 1; and can be found with this article online.

## Author Contributions

This is a single author contribution. All elements of this work, including the conception of ideas, formulation, and development of concepts, execution of experiments, analysis of data, and writing of the paper, were performed by the author.

## Acknowledgements

This work was made possible by the open public access policies of major grant funding agencies and international genomic databases and the willingness of many investigators worldwide to share their primary research data. I thank you Dr. Joshua Akey for the insightful comments during the preparation of the final versions of the manuscript. I would like to thank my anonymous colleagues for their valuable critical contributions during the peer review process of this work.

**Figure legends**

**Figure 1.** Two distinct pathways of human regulatory DNA divergence during evolution of human-specific genomic regulatory networks. Sequence conservation analyses of 18,364 candidate HSRS revealed two distinct patterns of regulatory DNA alignments to genomes of NHP: i) an alignment pattern with a significant majority (from 77.1% to 82.6%) of candidate HSRS being highly conserved in genomes of Bonobo and Chimpanzee (blue colored features in the figure); ii) an alignment pattern with only a minority (from 7.3% to 15.9%) of candidate HSRS being highly conserved in genomes of Bonobo and Chimpanzee (red colored features in the figure). It is proposed that these two distinct sequence conservation patterns reflect two mechanistically distinct pathways of human regulatory DNA divergence during evolution (see text for details). For each family of HSRS the percentage of highly conserved in NHP (blue) and human-specific (red) regulatory DNA segments are shown. The figure represents the graphical summary of the primary data reported in the Tables 7 and 12.

**Table 1.** Distribution of primate-specific and human-specific regulatory sequences among 524 haDHSs reported by Gittelman et al. (2015).

| Genomes/LiftOver setting | MinMatch 0.95 | MinMatch 0.99 | MinMatch 1.00 |
|---|---|---|---|
| Human genome (hg19) | 524 | 524 | 524 |
| Human genome (hg38) | 524 | 524 | 524 |
| Mouse genome conversion (mm10) | 298 | 148 | 66 |
| Percent conserved in rodents' genome | 56.9 | 28.2 | 12.6 |
| Chimpanzee genome conversion | 520 | 493 | 439 |
| Percent conserved in Chimpanzee | 99.2 | 94.1 | 83.8 |
| Chimpanzee conversion failures* | 4 | 31 | 85 |
| Bonobo genome conversion | 517 | 492 | 425 |
| Percent conserved in Bonobo | 98.7 | 93.9 | 81.1 |
| Bonobo conversion failures | 7 | 30 | 99 |
| Human-specific sequences** | 1 | 21 | 73 |
| Percent conserved in non-human primates | 99.8 | 96.0 | 86.1 |
| Percent of human-specific sequences | 0.2 | 4.0 | 13.9 |

Legends: LiftOver algorithm MinMatch, Minimum ratio of bases that must remap; haDHS, human accelerated DNase I hypersensitive sites;
*Chimpanzee genome PanTro4 conversion;
**Human-specific regulatory sequences were defined based on conversion failures to both Chimpanzee and Bonobo genomes

**Table 2.** Distribution of primate-specific and human-specific regulatory sequences among 2,741 HARs reported by Capra et al. (2013).

| Genomes/LiftOver setting | MinMatch 0.95 | MinMatch 0.99 | MinMatch 1.00 |
|---|---|---|---|
| Human genome (hg19) | 2,745 | 2,745 | 2,745 |
| Human genome (hg38) | 2,741 | 2,740 | 2,739 |
| Mouse genome conversion (mm10) | 2,364 | 1,642 | 1,004 |
| Percent conserved in rodents' genome | 86.2 | 59.9 | 36.7 |
| Chimpanzee genome conversion | 2,698 | 2,608 | 2,404 |
| Percent conserved in Chimpanzee | 98.4 | 95.2 | 87.8 |
| Chimpanzee conversion failures* | 43 | 133 | 337 |
| Bonobo genome conversion | 2,687 | 2,590 | 2,341 |
| Percent conserved in Bonobo | 98.0 | 94.5 | 85.5 |
| Bonobo conversion failures | 54 | 151 | 400 |
| Human-specific sequences** | 33 | 107 | 255 |
| Percent conserved in non-human primates | 98.8 | 96.1 | 90.7 |
| Percent of human-specific sequences | 1.2 | 3.9 | 9.3 |

Legends: LiftOver algorithm MinMatch, Minimum ratio of bases that must remap; HARs, human accelerated regions;
*Chimpanzee genome PanTro4 conversion;
**Human-specific regulatory sequences were defined based on conversion failures to both Chimpanzee and Bonobo genomes

**Table 3.** Distribution of primate-specific and human-specific regulatory sequences among 2,118 DHS fixed human specific regulatory regions reported by Marnetto et al. (2014).

| Genomes/LiftOver setting | MinMatch 0.95 | MinMatch 0.99 | MinMatch 1.00 |
|---|---|---|---|
| Human genome (hg19) | 2,118 | 2,118 | 2,118 |
| Human genome (hg38) | 2,116 | 2,115 | 2,114 |
| Mouse genome conversion (mm10) | 18 | 18 | 4 |
| Percent conserved in rodents' genome | 0.9 | 0.9 | 0.2 |
| Chimpanzee genome conversion | 5 | 5 | 5 |
| Percent conserved in Chimpanzee | 0.2 | 0.2 | 0.2 |
| Chimpanzee conversion failures* | 2,111 | 2,111 | 2,111 |
| Bonobo genome conversion | 375 | 331 | 242 |
| Percent conserved in Bonobo | 17.7 | 15.7 | 11.4 |
| Bonobo conversion failures | 1,741 | 1,785 | 1,874 |
| **Human-specific sequences**** | **1,737** | **1,781** | **1,869** |
| **Percent conserved in non-human primates** | **17.9** | **15.8** | **11.6** |
| **Percent of human-specific sequences** | **82.1** | **84.2** | **88.4** |

Legends: LiftOver algorithm MinMatch, Minimum ratio of bases that must remap; DHS, DNase I hypersensitive sites;
*Chimpanzee genome PanTro4 conversion;
**Human-specific regulatory sequences were defined based on conversion failures to both Chimpanzee and Bonobo genomes

**Table 4.** Distribution of primate-specific and human-specific regulatory sequences among 3,803 human specific transcription factor-binding sites reported by Glinsky (2015).

| Genomes/LiftOver setting | MinMatch 0.95 | MinMatch 0.99 | MinMatch 1.00 |
|---|---|---|---|
| Human genome (hg19) | 3,803 | 3,803 | 3,803 |
| Human genome (hg38) | 3,719 | 3,714 | 3,714 |
| Mouse genome conversion (mm10) | 31 | 13 | 12 |
| Percent conserved in rodents' genome | 0.8 | 0.4 | 0.3 |
| Chimpanzee genome conversion | 70 | 60 | 56 |
| Percent conserved in Chimpanzee | 1.9 | 1.6 | 1.5 |
| Chimpanzee conversion failures* | 3,649 | 3,659 | 3,663 |
| Bonobo genome conversion | 768 | 529 | 495 |
| Percent conserved in Bonobo | 20.7 | 14.2 | 13.3 |
| Bonobo conversion failures | 2,951 | 3,190 | 3,224 |
| **Human-specific sequences**** | **2,937** | **3,173** | **3,211** |
| **Percent conserved in non-human primates** | **21.0** | **14.6** | **13.5** |
| **Percent of human-specific sequences** | **79.0** | **85.4** | **86.5** |

Legends: LiftOver algorithm MinMatch, Minimum ratio of bases that must remap;
*Chimpanzee genome PanTro4 conversion;
**Human-specific regulatory sequences were defined based on conversion failures to both Chimpanzee and Bonobo genomes

**Table 5.** Distribution of primate-specific and human-specific regulatory sequences among 1,932 hESC fixed human specific regulatory regions reported by Marnetto et al. (2014).

| Genomes/LiftOver setting | MinMatch 0.95 | MinMatch 0.99 | MinMatch 1.00 |
|---|---|---|---|
| Human genome (hg19) | 1,932 | 1,932 | 1,932 |
| Human genome (hg38) | 1,932 | 1,930 | 1,928 |
| Mouse genome conversion (mm10) | 0 | 0 | 0 |
| Percent conserved in rodents' genome | 0.0 | 0.0 | 0.0 |
| Chimpanzee genome conversion | 1 | 1 | 0 |
| Percent conserved in Chimpanzee | 0.1 | 0.1 | 0.0 |
| Chimpanzee conversion failures* | 1,931 | 1,931 | 1,932 |
| Bonobo genome conversion | 575 | 529 | 396 |
| Percent conserved in Bonobo | 29.8 | 27.4 | 20.5 |
| Bonobo conversion failures | 1,357 | 1,403 | 1,536 |
| **Human-specific sequences**** | **1,357** | **1,403** | **1,536** |
| **Percent conserved in non-human primates** | **29.8** | **27.3** | **20.3** |
| **Percent of human-specific sequences** | **70.2** | **72.7** | **79.7** |

Legends: LiftOver algorithm MinMatch, Minimum ratio of bases that must remap; hESC, human embryonic stem cells;
*Chimpanzee genome PanTro4 conversion;
**Human-specific regulatory sequences were defined based on conversion failures to both Chimpanzee and Bonobo genomes

**Table 6.** Distribution of primate-specific and human-specific regulatory sequences among 4,249 fixed human specific regulatory regions reported by Marnetto et al. (2014).

| Genomes/LiftOver setting | MinMatch 0.95 | MinMatch 0.99 | MinMatch 1.00 |
|---|---|---|---|
| Human genome (hg19) | 4,249 | 4,249 | 4,249 |
| Human genome (hg38) | 4,249 | 4,248 | 4,235 |
| Mouse genome conversion (mm10) | 0 | 0 | 0 |
| Percent conserved in rodents' genome | 0.0 | 0.0 | 0.0 |
| Chimpanzee genome conversion | 23 | 21 | 13 |
| Percent conserved in Chimpanzee | 0.5 | 0.5 | 0.3 |
| Chimpanzee conversion failures* | 4,226 | 4,228 | 4,232 |
| Bonobo genome conversion | 653 | 590 | 438 |
| Percent conserved in Bonobo | 15.4 | 13.9 | 10.3 |
| Bonobo conversion failures | 3,596 | 3,659 | 3,811 |
| Human-specific sequences** | 3,580 | 3,645 | 3,800 |
| Percent conserved in non-human primates | 15.7 | 14.2 | 10.3 |
| Percent of human-specific sequences | 84.3 | 85.8 | 89.7 |

Legends: LiftOver algorithm MinMatch, Minimum ratio of bases that must remap; hESC, human embryonic stem cells;
*Chimpanzee genome PanTro4 conversion;
**Human-specific regulatory sequences were defined based on conversion failures to both Chimpanzee and Bonobo genomes.

**Table 7.** Distribution of highly conserved in non-human primates regulatory sequences among 15,371 candidate human-specific regulatory sequence.

| HSRS/Genomes | haDHS | HARs | HSTFBS | DHS_FHSRR | hESC_FHSRR | Other_FHSRR |
|---|---|---|---|---|---|---|
| **Human genome (hg19)** | 524 | 2,745 | 3,803 | 2,118 | 1,932 | 4,249 |
| **Human genome (hg38)** | 524 | 2,739 | 3,714 | 2,114 | 1,928 | 4,235 |
| **Mouse genome conversion (mm10)** | 66 | 1,004 | 12 | 4 | 0 | 0 |
| **Reciprocal conversion to human genome** | 23 | 560 | 1 | 2 | 0 | 0 |
| **Percent conserved in rodents' genome** | **4.4** | **20.4** | **0.0** | **0.1** | **0.0** | **0.0** |
| **Chimpanzee genome conversion** | 439 | 2,404 | 56 | 5 | 0 | 13 |
| **Reciprocal conversion to human genome** | 390 | 2,146 | 40 | 0 | 0 | 1 |
| **Percent conserved in Chimpanzee** | 74.4 | 78.3 | 1.1 | 0 | 0 | 0 |
| **Bonobo genome conversion** | 425 | 2,341 | 495 | 242 | 396 | 438 |
| **Reciprocal conversion to human genome** | 383 | 2,123 | 262 | 199 | 306 | 350 |
| **Percent conserved in Bonobo** | 73.1 | 77.5 | 7.1 | 9.4 | 15.9 | 8.3 |
| **Conserved in non-human primates**\*\* | **404** | **2,262** | **271** | **199** | **306** | **351** |
| **Percent conserved in non-human primates** | **77.1** | **82.6** | **7.3** | **9.4** | **15.9** | **8.3** |
| **Bonobo & Chimp conserved** | 370 | 2,004 | 31 | 0 | 0 | 0 |
| **Chimp only conserved** | 21 | 141 | 9 | 0 | 0 | 1 |
| **Bonobo only conserved** | 13 | 117 | 231 | 199 | 306 | 350 |

Legends: LiftOver algorithm MinMatch Minimum ratio of bases that must remap) threshold was 1.00
\*Chimpanzee genome PanTro4 conversion;
\*\*Conserved in non-human primates sequences were defined based on both direct and reciprocal conversions to either one or both Chimpanzee and Bonobo genomes at MinMatch threshold of 1.00;
HSRS, human-specific regulatory sequences;
HSTFBS, human-specific transcription factor-binding sites;
haDHS, human accelerated DNase I hypersensitive sites;
HARs, human accelerated regions;
DHS, DNase I hypersensitive sites;
FHSRR, fixed human-specific regulatory regions;

**Table 8.** Distribution of primate-specific and human-specific regulatory sequences among 1,000 human-biased divergent cranial neural crest cells' enhancers reported by Prescott et al. (2015).

| Genomes/LiftOver setting | MinMatch 0.95 | MinMatch 0.99 | MinMatch 1.00 |
|---|---|---|---|
| Human genome (hg19) | 1,000 | 1,000 | 1,000 |
| Human genome (hg38) | 996 | 994 | 991 |
| Mouse genome conversion (mm10) | 201 | 62 | 21 |
| Percent conserved in rodents' genome | 20.2 | 6.2 | 2.1 |
| Chimpanzee genome conversion | 976 | 943 | 871 |
| Percent conserved in Chimpanzee | 97.99 | 94.7 | 87.4 |
| Chimpanzee conversion failures* | 20 | 53 | 125 |
| Bonobo genome conversion | 957 | 927 | 844 |
| Percent conserved in Bonobo | 96.1 | 93.1 | 84.7 |
| Bonobo conversion failures | 39 | 69 | 152 |
| **Human-specific sequences*** | **17** | **42** | **106** |
| **Percent conserved in non-human primates** | **98.3** | **95.8** | **89.4** |
| **Percent of human-specific sequences** | **1.7** | **4.2** | **10.6** |

Legends: LiftOver algorithm MinMatch, Minimum ratio of bases that must remap; haDHS, human accelerated DNase I hypersensitive sites;
*Chimpanzee genome PanTro4 conversion;
**Human-specific regulatory sequences were defined based on conversion failures to both Chimpanzee and Bonobo genomes

**Table 9.** Distribution of primate-specific and human-specific regulatory sequences among 1,000 chimp-biased divergent cranial neural crest cells' enhancers reported by Prescott et al. (2015).

| Genomes/LiftOver setting | MinMatch 0.95 | MinMatch 0.99 | MinMatch 1.00 |
|---|---|---|---|
| Human genome (hg19) | 1,000 | 1,000 | 1,000 |
| Human genome (hg38) | 999 | 999 | 998 |
| Mouse genome conversion (mm10) | 212 | 86 | 30 |
| Percent conserved in rodents' genome | 21.2 | 8.6 | 3.0 |
| Chimpanzee genome conversion | 977 | 947 | 884 |
| Percent conserved in Chimpanzee | 97.9 | 94.8 | 88.5 |
| Chimpanzee conversion failures* | 22 | 52 | 115 |
| Bonobo genome conversion | 956 | 917 | 847 |
| Percent conserved in Bonobo | 95.7 | 91.8 | 84.8 |
| Bonobo conversion failures | 43 | 82 | 152 |
| **Human-specific sequences*** | **13** | **38** | **85** |
| **Percent conserved in non-human primates** | **98.7** | **96.1** | **91.5** |
| **Percent of human-specific sequences** | **1.3** | **3.8** | **8.5** |

Legends: LiftOver algorithm MinMatch, Minimum ratio of bases that must remap; haDHS, human accelerated DNase I hypersensitive sites;
*Chimpanzee genome PanTro4 conversion;
**Human-specific regulatory sequences were defined based on conversion failures to both Chimpanzee and Bonobo genomes

**Table 10.** Distribution of primate-specific and human-specific regulatory sequences among 410 human-specific H3K4me3 histone signatures at TSS* divergent in human prefrontal cortex reported by Shulha et al. (2012).

| Genomes/LiftOver setting | MinMatch 0.95 | MinMatch 0.99 | MinMatch 1.00 |
|---|---|---|---|
| Human genome (hg19) | 410 | 410 | 410 |
| Human genome (hg38) | 406 | 401 | 394 |
| Mouse genome conversion (mm10) | 8 | 0 | 0 |
| Percent conserved in rodents' genome | 1.97 | 0 | 0 |
| Chimpanzee genome conversion | 298 | 235 | 86 |
| Percent conserved in Chimpanzee | 73.4 | 57.9 | 21.2 |
| Chimpanzee conversion failures** | 108 | 171 | 320 |
| Bonobo genome conversion | 263 | 202 | 74 |
| Percent conserved in Bonobo | 64.8 | 49.8 | 18.2 |
| Bonobo conversion failures | 143 | 204 | 332 |
| **Human-specific sequences*** | **72** | **131** | **299** |
| **Percent conserved in non-human primates** | **82.3** | **67.7** | **26.4** |
| **Percent of human-specific sequences** | **17.7** | **32.3** | **73.6** |

Legends: LiftOver algorithm MinMatch, Minimum ratio of bases that must remap; haDHS, human accelerated DNase I hypersensitive sites; *TSS, transcription start sites;
**Chimpanzee genome PanTro4 conversion;
***Human-specific regulatory sequences were defined based on conversion failures to both Chimpanzee and Bonobo genomes

**Table 11.** Distribution of primate-specific and human-specific regulatory sequences among 583 regions of human-specific deletions of regulatory DNA (hCONDELs) reported by McLean et al. (2011).

| Genomes/LiftOver setting | MinMatch 0.95 | MinMatch 0.99 | MinMatch 1.00 |
|---|---|---|---|
| Human genome (hg18) | 583 | 583 | 583 |
| Human genome (hg38) | 246 | 246 | 245 |
| Mouse genome conversion (mm10) | 23 | 22 | 22 |
| Percent conserved in rodents' genome | 9.3 | 8.9 | 8.9 |
| Chimpanzee genome conversion | 28 | 24 | 17 |
| Percent conserved in Chimpanzee | 11.4 | 9.8 | 6.9 |
| Chimpanzee conversion failures* | 218 | 222 | 229 |
| Bonobo genome conversion | 101 | 94 | 71 |
| Percent conserved in Bonobo | 41.1 | 38.2 | 28.9 |
| Bonobo conversion failures | 145 | 152 | 175 |
| **Human-specific sequences**** | **140** | **147** | **168** |
| **Percent conserved in non-human primates** | **43.1** | **40.2** | **31.7** |
| **Percent of human-specific sequences** | **56.9** | **59.8** | **68.3** |

Legends: LiftOver algorithm MinMatch, Minimum ratio of bases that must remap; hESC, human embryonic stem cells;
*Chimpanzee genome PanTro4 conversion;
**Human-specific regulatory sequences were defined based on conversion failures to both Chimpanzee and Bonobo genomes

Table 12. Distribution of highly conserved in non-human primates regulatory sequences among candidate human-specific regulatory sequence defined by the functional divergence from chimpanzee or deletions of ancestral DNA in the human genome

| HSRS/Genomes | Human-biased CNCC's enhancers | Chimp-biased CNCC's enhancers | hCONDELs | H3K4me3 signatures in human prefrontal neurons | All HSRS |
|---|---|---|---|---|---|
| **Human genome (hg19)** | 1,000 | 1,000 | 583 | 410 | 18,364 |
| **Human genome (hg38)** | 996 | 998 | 245 | 394 | 17,887 |
| **Mouse genome conversion (mm10)** | 21 | 30 | 22 | 0 | 1,159 |
| **Reciprocal conversion to human genome** | 4 | 7 | 18 | 0 | 615 |
| **Percent conserved in rodents' genome** | 0.4 | 0.7 | 7.3 | 0 | 3.4 |
| **Chimpanzee genome conversion** | 871 | 884 | 17 | 86 | 4,775 |
| **Reciprocal conversion to human genome** | 765 | 785 | 12 | 36 | 4,175 |
| **Percent conserved in Chimpanzee** | 76.8 | 78.7 | 4.9 | 9.1 | 23.3 |
| **Bonobo genome conversion** | 844 | 847 | 71 | 74 | 6,173 |
| **Reciprocal conversion to human genome** | 754 | 760 | 63 | 36 | 5,236 |
| **Percent conserved in Bonobo** | 75.7 | 76.2 | 25.7 | 9.1 | 29.3 |
| **Conserved in non-human primates**** | **804** | **820** | **68** | **50** | **5,535** |
| **Percent conserved in non-human primates** | **80.7** | **82.2** | **27.8** | **12.7** | **30.9** |
| **Bonobo & Chimp conserved** | 715 | 725 | 7 | 22 | 3,874 |
| **Chimp only conserved** | 50 | 60 | 5 | 14 | 301 |
| **Bonobo only conserved** | 39 | 35 | 56 | 14 | 1,360 |

Legends: LiftOver algorithm MinMatch Minimum ratio of bases that must remap) threshold was 1.00
*Chimpanzee genome PanTro4 conversion;
**Conserved in non-human primates sequences were defined based on both direct and reciprocal conversions to either one of both Chimpanzee and Bonobo genomes at MinMatch threshold of 1.00;
HSRS, human-specific regulatory sequences;
hCONDELs, human-specific deletions of regulatory DNA;
CNCCs, cranial neural crest cells;
All HSRS column shows the sum of records for each categories from the corresponding entries in Tables 7 & 12.

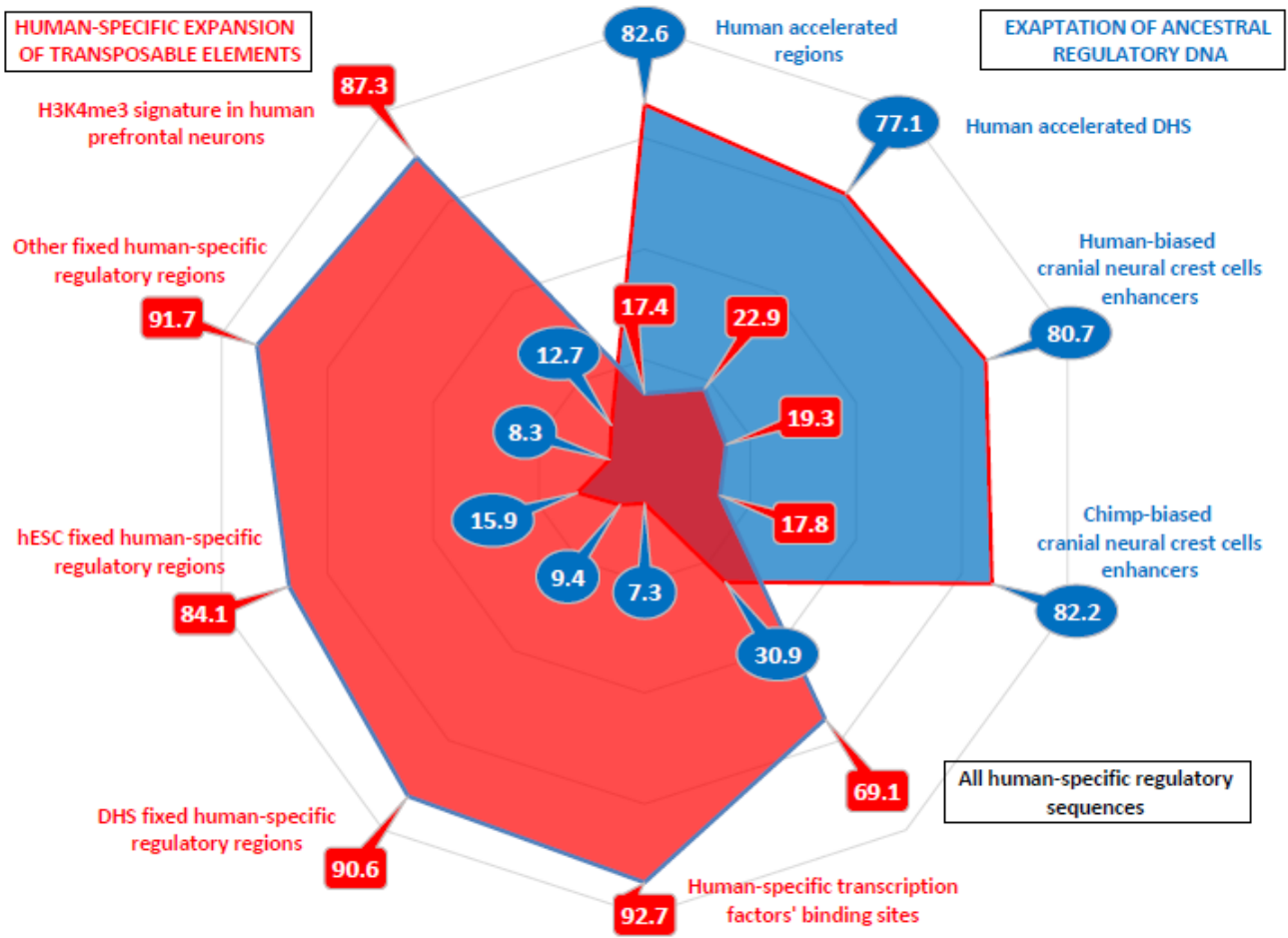